\documentclass[11pt]{article}
\pdfoutput=1

\usepackage{geometry}
\usepackage{slashed}
\usepackage{graphicx}
\usepackage{amsmath}
\usepackage{amssymb}
\usepackage{epstopdf}
\usepackage{verbatim}
\usepackage{url}
\usepackage{amsfonts}
\usepackage{layout}
\usepackage{slashed}
\usepackage[usenames,dvipsnames]{color}

\usepackage{hyperref}	% This is for pdftex

\fontsize{15}{15}

\numberwithin{equation}{section}

\newcommand{\be}{\begin{equation}}
\newcommand{\bea}{\begin{eqnarray}}
\newcommand{\beq}[1]{\begin{equation}\label{#1}}

\newcommand{\ee}{\end{equation}}
\newcommand{\eea}{\end{eqnarray}}
\newcommand{\eeq}{\end{equation}}
\newcommand{\bi}{\begin{itemize}}
\newcommand{\ei}{\end{itemize}}
\newcommand{\lsim}{\!\mathrel{\hbox{\rlap{\lower.55ex \hbox{$\sim$}} \kern-.34em \raise.4ex \hbox{$<$}}}}
\newcommand{\gsim}{\!\mathrel{\hbox{\rlap{\lower.55ex \hbox{$\sim$}} \kern-.34em \raise.4ex \hbox{$>$}}}}
\newcommand{\mr}[1]{\mathrm{#1}}

\newcommand{\vev}[1]{\langle #1 \rangle}
\newcommand{\st}{$^\mathrm{st}$\,}

\begin{document}

\begin{titlepage}

\begin{flushright}
MCTP-11-35 \\
SLAC-PUB-14621\\
\end{flushright}
\vspace{3.0cm}

\begin{center}
{\Large \bf Electroweak Baryogenesis and Colored Scalars}

\vspace{0.2in}
{\bf Timothy Cohen$^{a,b}$ and Aaron Pierce$^{b}$}\\
\vspace{0.2cm}
{\it $^a$SLAC National Accelerator Laboratory,\\
2575 Sand Hill Rd, Menlo Park, CA 94025}\\
\vspace{0.2cm}
{\it $^b$Michigan Center for Theoretical Physics (MCTP), \\
Department of Physics, University of Michigan, Ann Arbor, MI
48109}\\
\end{center}

\vspace{0.4cm}

\begin{abstract}
We consider the 2-loop finite temperature effective potential for a Standard Model-like Higgs boson, allowing Higgs boson couplings to additional scalars.   If the scalars transform under color, they contribute 2-loop diagrams to the effective potential that include gluons.  These 2-loop effects are perhaps stronger than previously appreciated.  For a Higgs boson mass of 115 GeV, they can increase the strength of the phase transition by as much as a factor of 3.5.  It is the analogue of this effect that is responsible for the survival of the tenuous electroweak baryogenesis window of the Minimal Supersymmetric Standard Model.  We further illuminate the importance of these 2-loop diagrams by contrasting models with colored scalars to models with singlet scalars.  We conclude that baryogenesis favors models with light colored scalars.  This motivates searches for  pair-produced di-jet resonances or jet(s) + $\slashed{E}{}_T$.  
\end{abstract}

\end{titlepage}

%\tableofcontents

\newpage

\setcounter{equation}{0}
%%%%%%%%%%%%%%%%%%%%%%%%%%%%%%%%%%%%%%%%%%%%%%%%%%%%%%%%%%%%%%%%%%%%%%%%%%%%

\section{Introduction}
Electroweak baryogenesis generates the baryon asymmetry of the universe via the physics of the weak scale.  For reviews, see \cite{Cohen:1993nk,Trodden:1998ym}.  These models satisfy the out-of-equilibrium Sakharov criterion  with a 1\st order electroweak phase transition --- as bubbles of true vacuum percolate, they separate regions with full strength sphaleron reactions from those where these effects are suppressed.  This prevents wash out of the generated asymmetry.  
1\st order phase transitions have a critical temperature $T_C$, where the potential exhibits two degenerate minima, one at zero and the other at $\phi_C$, the critical value for the background field $\phi$.  We are interested in models which yield $\phi_C/T_C > 0.9$, in order to avoid sphaleron washout effects \cite{Carena:2008vj}.  While a phase transition with the appropriate strength is in principle realizable with Standard Model particle content, the bound on the Higgs boson mass from LEP eliminates this possibility (for a review see \cite{Quiros:1999jp}).  

What modifications beyond the Standard Model allow the phase transition to be more strongly 1\st order?  One possibility is that the structure of the phase transition is more complicated, \emph{i.e.}, there exist other fields which obtain vacuum expectation values (vevs). Another approach, taken here, is to assume a Standard Model-like Higgs boson alone obtains a vev, but couplings of new states to the Higgs boson strengthen this phase transition.  A potential example is the Minimal Supersymmetric Standard Model (MSSM) in the decoupling limit, where there is a large coupling between the top squark and the Higgs boson.  However, if a 1-loop analysis is performed,  the collider bounds on the stop and Higgs boson masses indicate a phase transition that is not strong enough.  If instead one performs the calculation at two loops \cite{Espinosa:1996qw, Carena:1997ki}, a small window of parameter space survives (where the soft mass for the stop squarks is as negative as possible while avoiding tunneling to a charge-color breaking vacuum) \cite{Carena:2008vj}.  This enhancement was first emphasized in \cite{Espinosa:1996qw}.  We reexamine the importance of these 2-loop effects and find an increase in the 1\st orderness from one loop to two loops (as measured by $\phi_{C}/T_{C}$) by as much as a factor of 3.5.  This can be contrasted with \cite{Espinosa:1996qw} where, when one restricts the MSSM parameter space to only include positive values of the bare stop mass, a maximum increase of 75\% was observed  (see Fig.~\ref{fig:VariousContributionsToPotential} below for details on how the enhancement depends on this parameter).  There are good reasons to expect the 2-loop effects to be anomalously large, and we expect 3-loop effects to be under control.  We discuss this in more detail below. 

In this paper, we investigate the electroweak phase transition for models containing new scalars with different quantum numbers and arbitrary coupling strengths to the Higgs boson.\footnote{Another necessary ingredient for successful electroweak baryogenesis is the presence of new CP violation near the weak scale.  We remain agnostic about the origin of this novel CP violation.  We assume this additional physics has a subdominant impact on the dynamics of the phase transition.  This is true in many cases, for example MSSM based models which rely on a phase between $\mu$ and gaugino masses.}    Does this freedom make it much easier to achieve a 1\st order phase transition?  As in the MSSM, 2-loop effects can be important.  To appreciate the impact of the 2-loop effects (and the color quantum numbers of the new scalars), we consider two simple extensions of the Standard Model:
\begin{enumerate}
\item A colored fundamental complex scalar $X_c$, that couples to the Higgs boson with an arbitrary quartic; and
\item A sextuplet of uncolored scalars $X_0$, that couple to the Higgs boson with an arbitrary quartic.  We select a toy with six degrees of freedom to match the previous case.  This helps to isolate the importance of the color quantum numbers for the phase transition. 
\end{enumerate}

One might think that considering phase transitions with a single additional scalar is an overly restrictive set-up.  However, only light states with mass $\lesssim v$ will be a part of the thermal bath at temperatures relevant to the phase transition.  Since the LHC has not discovered a plethora of new light states, it is plausible that the sector that modifies the dynamics of the electroweak phase transition can be described by a single field.  In this case, our analysis captures the relevant physics  (re-emphasizing that we concentrate on the possibility that there is a single phase transition near the weak scale).

In Sec.~\ref{sec:EWPTat2Loops}, we discuss how the $SU(3)$ quantum numbers play an important role in making the phase transition more strongly 1\st order.  Since it should be possible to produce colored scalars at a hadron collider, this motivates a connection with LHC phenomenology, which is the focus of Sec.~\ref{sec:ColliderPheno}.  

\section{The Electroweak Phase Transition at 2 Loops}\label{sec:EWPTat2Loops}
\begin{figure}[h]
\begin{centering}
\vspace{0.4cm}
\includegraphics[width=.7\textwidth]{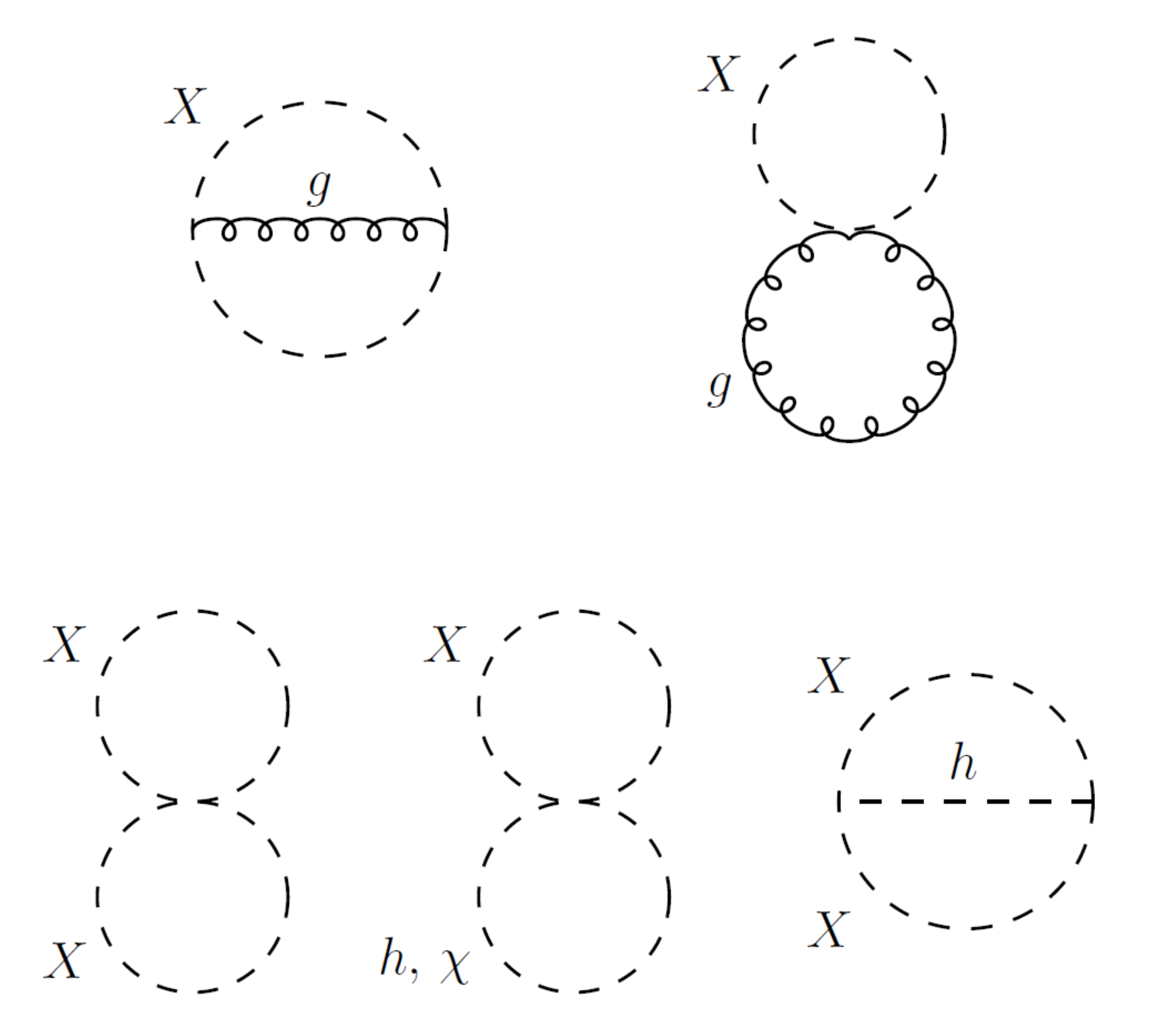}
\vspace{0.2cm}
\caption{\label{fig:2LoopDiagrams} \small The dominant 2-loop diagrams involving a new scalar state, $X$, which couples to the Higgs boson.  Dashed lines are scalars and curly lines are gluons.  $h$ is the Higgs boson, and $\chi$ is the Goldstone boson.  Note that if $X$ is a singlet the diagrams involving $X$ and gluons do not exist.}
\end{centering}
\end{figure}

Before presenting specific models, we briefly discuss some generic features of the electroweak phase transition at two  loops.  It was recognized in \cite{Dine:1992vs} that 2-loop effects can have a qualitative impact on the strength of the electroweak phase transition.  At the critical temperature, the 2-loop contributions to the potential introduce new terms which effectively contribute to the mass of $\phi$, and (assuming they are positive) act to maintain $\phi = 0$ to lower temperatures.  These new contributions postpone the phase transition. As we will discuss below, this has the effect of making it more strongly 1\st order.  For models that couple a new scalar, $X$, to the Higgs boson, the leading 2-loop diagrams are shown in Fig.~\ref{fig:2LoopDiagrams}.  (We neglect new physics diagrams proportional to $(g^{\prime})^2$.)  There are also 2-loop contributions that involve Standard Model states --- the dominant terms due to these interactions are given in \emph{e.g.}, \cite{Espinosa:1996qw}.

We are interested in models with negative bare mass squared parameters for the Higgs fields and for $X$.  This is raises a technical issue.  In the finite temperature potential, taking the high temperature expansion for bosons, there is a term $\sim T\, m^3(\phi)$ where $m(\phi)$ is the field-dependent mass of the relevant boson.  If this boson has a negative bare mass squared, this term exhibits non-analytic behavior for $\phi \rightarrow 0$.  There is also a breakdown in the loop expansion (coming from the IR contribution to loop momenta) as one approaches the critical temperature.  These two problems are solved simultaneously by performing a resummation of the propagators of the zero Matsubara modes,  incorporating the leading finite temperature corrections to their masses for all scalars.  (Similar techniques are applied for longitudinal gauge bosons.) This adds a positive temperature dependent contribution to the zero mode mass, thereby removing the problem of non-analyticity, \emph{i.e.}, $m_\mr{resummed}(\phi,T) > 0$ as $\phi \rightarrow 0$.  It also  cures the IR divergences.   At one loop, this is known as ``daisy resummation." Techniques for including these effects at two loops were described in detail in \cite{Arnold:1992rz}, and we follow their prescription for our evaluation of the diagrams in Fig.~\ref{fig:2LoopDiagrams}.

\subsection{The Electroweak Phase Transition with a New Colored Scalar}\label{sec:EWPTwithColoredScalar}
In this section we discuss the strength of the electroweak phase transition for a model with a new colored scalar in the fundamental representation  $X_c$,\footnote{In principle, gauge invariance would allow any representation of the colored scalar.  We choose the fundamental representation for simplicity.  We note that in the case of an adjoint which decays to a pair of jets, the LHC has already excluded the mass range from 100-185 (except for a 5 GeV window at 140 GeV) \cite{Zhu}.  See Sec.~\ref{sec:ColliderPheno} below for more details.} with hypercharge $1/3$ and a bare mass term $M_X$.  It couples to the Higgs boson via a quartic coupling, $Q$:
\be\label{eq:XLagrangian}
\mathcal{L} \supset -M_X^2 |X_c|^2 - Q |X_c|^2 |H|^2 - \frac{K}{6} |X_c|^4,
\ee 
where $K$ is the quartic self coupling for $X_c$, and $H$ is the Higgs doublet.  We follow the notation of the ``light stop effective theory" \cite{Carena:2008rt}.  Indeed, this model is essentially identical to that model (except for the lack of a top-stop-Higgsino coupling).  In addition, in the supersymmetric case, one must impose the relevant supersymmetric boundary conditions for $Q$ and $K$ at the scale where supersymmetry is restored.  For the MSSM in the baryogenesis window, where $M_X^2 = - (80 \mbox{ GeV})^2$, $0.3 \lesssim Q \lesssim 0.85$.  The range in $Q$ is largely determined by the choice of the stop trilinear $a$-term \cite{Menon:2009mz, Carena:2008rt}.\footnote{A model that can realize the freedom of essentially arbitrary $Q$ is the MSSM augmented by two sets of vector-like colored matter, married by non-zero couplings to the Higgs boson \cite{Martin:2009bg}, in the limit where the spectrum is taken such that only one of the light colored scalar states survives at low energies.} For our parameter scans, we take $K = 1.6 \simeq g_s^2+4/3\,g'^2$ since in supersymmetric models it would be dominated by the $D$-term contribution.  The renormalization group scale is taken to be $m_\mr{top}$.

\begin{figure}[t]
\begin{centering}
\includegraphics[width=.7\textwidth]{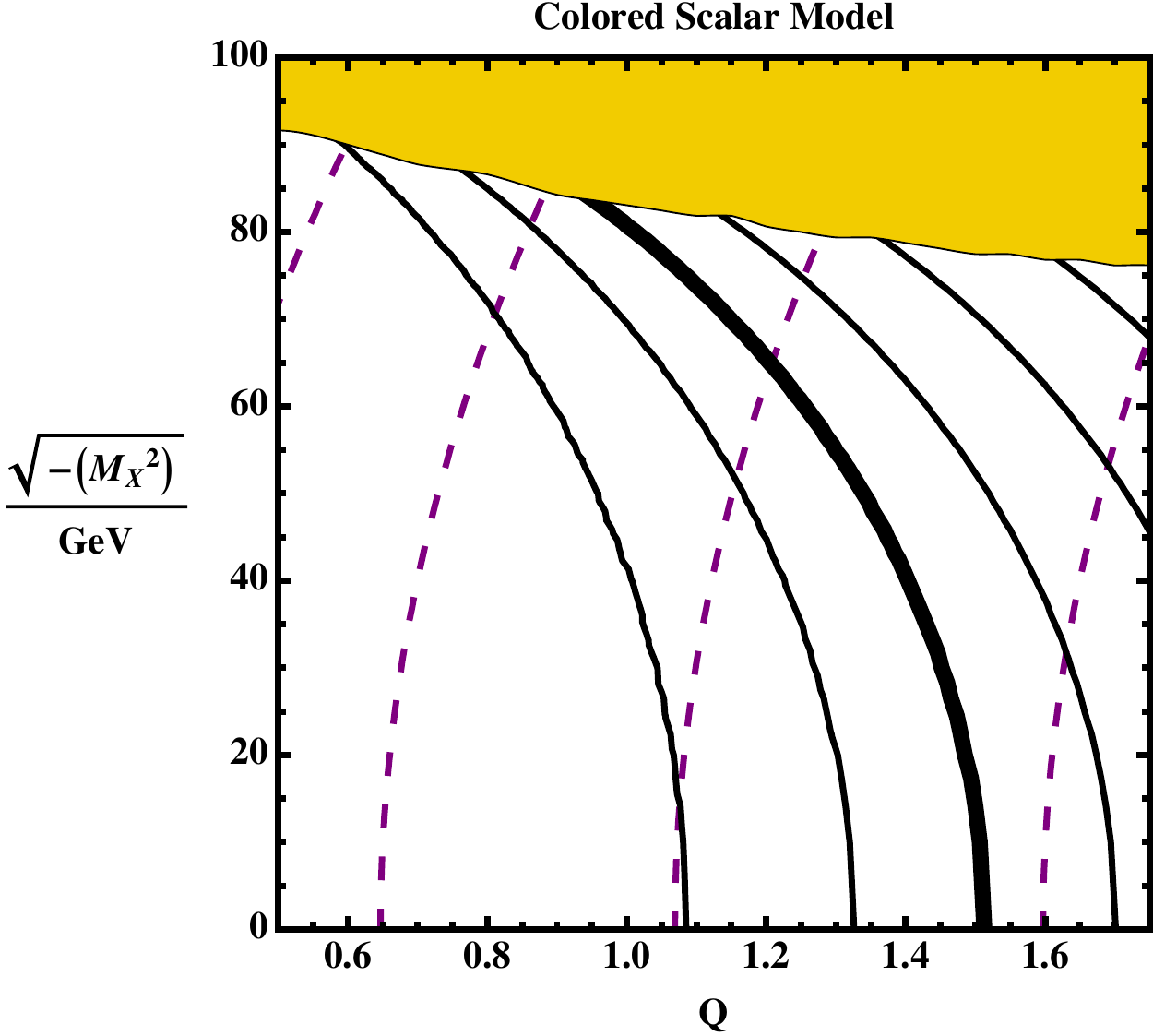}
\caption{\label{fig:ContourMagicNumber2Loop_Colored} \small Contours of $\phi_C/T_C$ [black, solid] and the physical $X$ mass [purple, dashed] in the $\sqrt{-M_X^2}$ vs.~$Q$ plane in the model with an additional colored scalar, $X_c$.  The Higgs mass is fixed to be 115 GeV and $K=1.6$.  The thickest solid contour marks $\phi_C/T_C = 0.9$ and the contours to the right (left) of it mark successive values of $+$ ($-$) 0.2.  The dashed contours are for $X$ masses of 140 GeV, 180 GeV, 220 GeV, and 260 GeV from left to right.  First orderness is maximized for larger values of $Q$ and more negative values of $M_X^2$.  In the yellow region there would be a phase transition to a vacuum with $\vev{X_c}\neq 0$. The maximum value of $Q$ plotted corresponds to where the high temperature expansion begins to break down.}
\end{centering}
\end{figure}

Using the Lagrangian of Eq.~(\ref{eq:XLagrangian}), we compute the strength of the electroweak phase transition.  The result of the resummed 2-loop computation is shown in Fig.~\ref{fig:ContourMagicNumber2Loop_Colored} where we have plotted contours of $\phi_C/T_C$ in the $\sqrt{-M_X^2}$ vs.~$Q$ plane.  We have fixed the Higgs mass to be $115 \mbox{ GeV}$ by adjusting the value of the Higgs quartic.  As expected,  the phase transition is strengthened as the $Q$ coupling is increased.    Also, for a fixed value of $Q$, going to more negative values of $M_X^2$ increases the 1\st orderness.  

The region with $M_X^2<0$ was studied in the MSSM using a 1-loop analysis in \cite{Carena:1996wj}.  For $Q > g_\mr{weak}$, the term which dominantly drives the strength of the phase transition at one loop is $V^\mr{1-loop} \supset -n_X \, T \overline{M}_X^3$/12 .  Here $n_X$ is the number of degrees of freedom of the $X_c$, and $\overline{M}_X$ is the resummed mass for $X_c$:
\be
\overline{M}_X^2 \equiv M_X^2+\frac{1}{2}\,Q\, \phi^2 + \Pi_c(T),
\ee
and
\be\label{eq:PI}
\Pi_c(T) =  \left( \frac{1}{3}\, g_3^2+\frac{5}{27}\, g'^2 + \frac{1}{9}\, K + \frac{1}{6}\, Q \right) T^2,
\ee
is the thermal mass of the $X_c$.  This term makes its maximum contribution to $\phi_C/T_C$ when $M_X^2 \simeq -\Pi_c(T_C)$ since in this limit $\overline{M}_X^3 \sim Q\, \phi^3$.  In the MSSM when $X_c$ is a stop, the stop-top-Higgsino Yukawa coupling also yields a contribution to $\Pi_c$.

Since $M_X^2 < 0$, the universe could conceivably evolve to a vacuum with $\vev{X_c}\neq0$ where color and electric charge is broken.  See \cite{Carena:1996wj} for a discussion in the context of the MSSM.  To ensure the electroweak phase transition proceeds first, the temperature for nucleating bubbles of the $\vev{X_c}\neq 0$ vacuum must be lower than that for nucleating bubbles of $\vev{\phi}\neq 0$ vacuum.  In \cite{Carena:2008vj} the bounce action was computed for the case of MSSM boundary conditions.  They conclude that the desired vacuum is achieved as long as $T_C \gtrsim (T_C)_X + 1.6 \mbox{ GeV}$, where $(T_C)_X$ is the 2-loop critical temperature for the $X_c$ direction.  We apply this constraint to exclude the yellow region in the $\sqrt{-M_X^2}$ vs.~$Q$ plane of Fig.~\ref{fig:ContourMagicNumber2Loop_Colored}.  Because we allow parameter space beyond that accessible in the MSSM, we do not expect this condition to precisely apply over the entire range of $Q$; we leave the study of the dependence of this constraint on the parameters for future work.  

\begin{figure}[t]
\begin{centering}
\includegraphics[width=.65\textwidth]{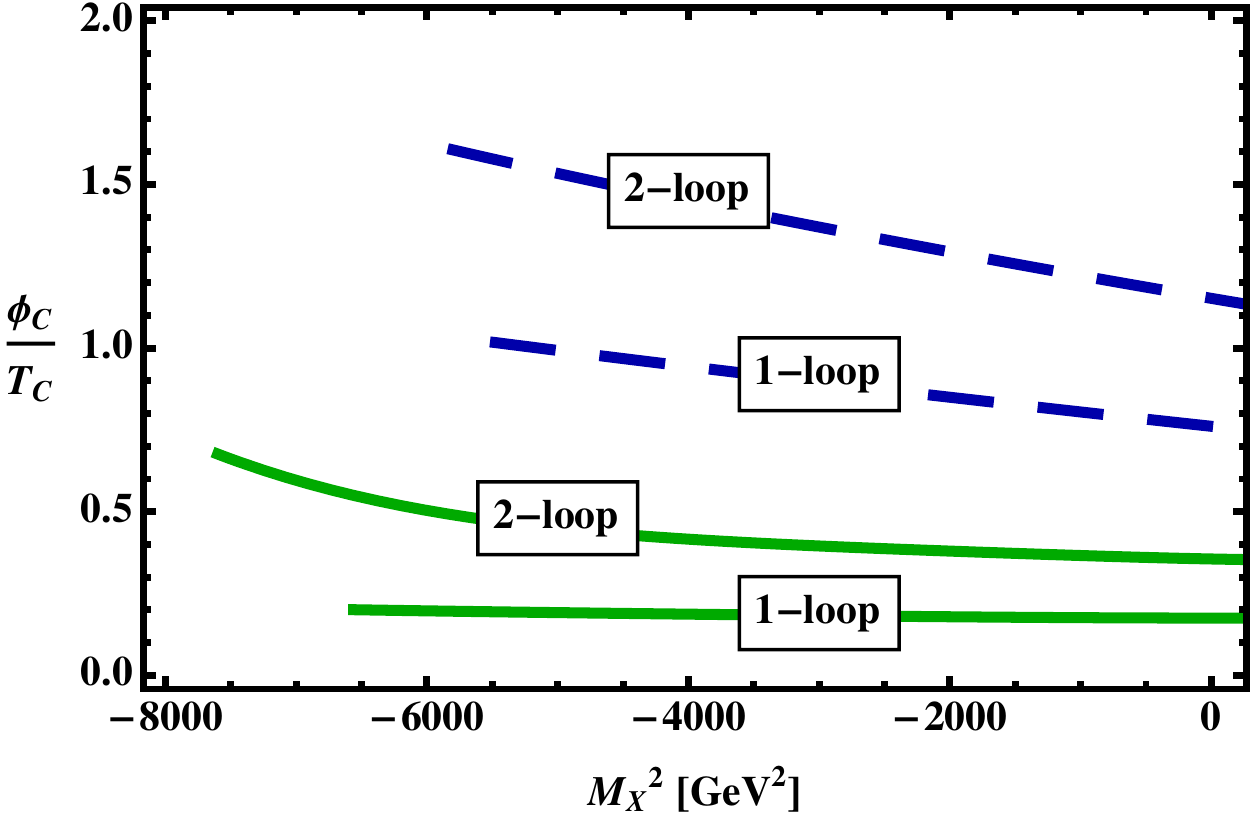}
\caption{\label{fig:Compare2LoopTo1LoopVsMUSq} \small The 2-loop and 1-loop values of $\phi_C/T_C$ as a function of $M_X^2$ for $Q = 0.75$ [green, solid] and $Q = 1.75$ [blue, dashed].  The other parameters are fixed to be $K = 1.6$ and $m_h = 115 \mbox{ GeV}$.  The minimum value of $M_X^2$ plotted corresponds to the boundary where the $\vev{X_c}\neq 0$ phase transition would occur before the Higgs phase transition.}
\end{centering}
\end{figure}

As $M_X^2$ becomes more negative, the importance of the 2-loop diagrams increases dramatically.  We illustrate this in Fig.~\ref{fig:Compare2LoopTo1LoopVsMUSq}, where we have plotted $\phi_C/T_C$ computed at two loops and at one loop as a function of $M_X^2$ and for two values of $Q$.  The 2-loop effects can increase the strength of the phase transition by as much as a factor of $\sim 3.5$.  This dramatic enhancement is not specific to our toy model, it is also present in the MSSM.  One should not dismay at the presence of this huge shift at two loops.  First, the coefficient of $\phi^2$ approximately cancels in the 1-loop analysis, leaving the 2-loop effects to dominate.  Second, there are new couplings that first appear at two loops, so the contribution is the leading one in these couplings.  For these reasons, we expect the 3-loop effects to be under control.  It would be interesting to calculate them, but that is beyond the scope of this work.  
 
\begin{figure}[t!]
\begin{centering}
$\begin{array}{ccc}
\includegraphics[width=.465\textwidth]{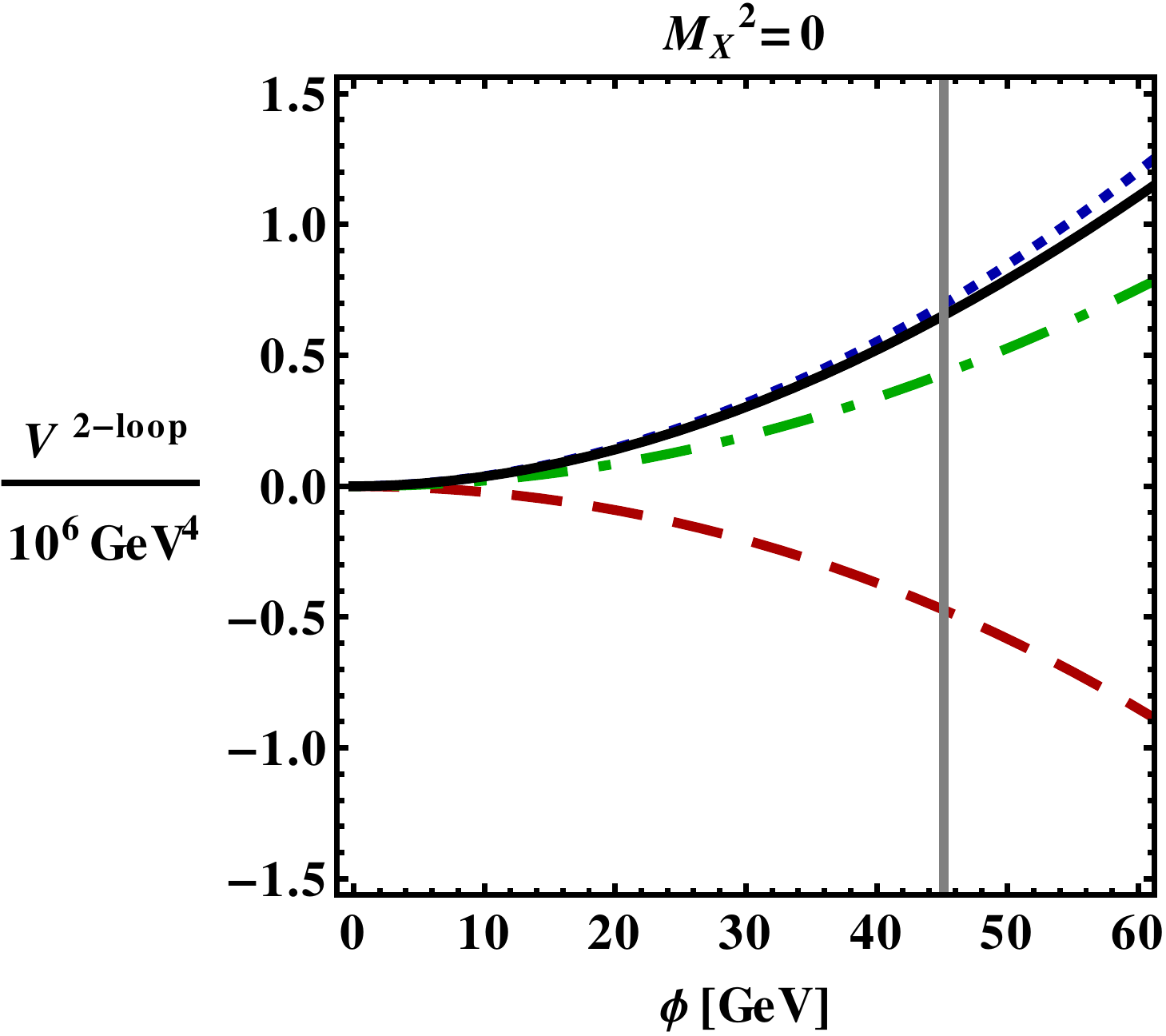} & $\quad$ & \includegraphics[width=.45\textwidth]{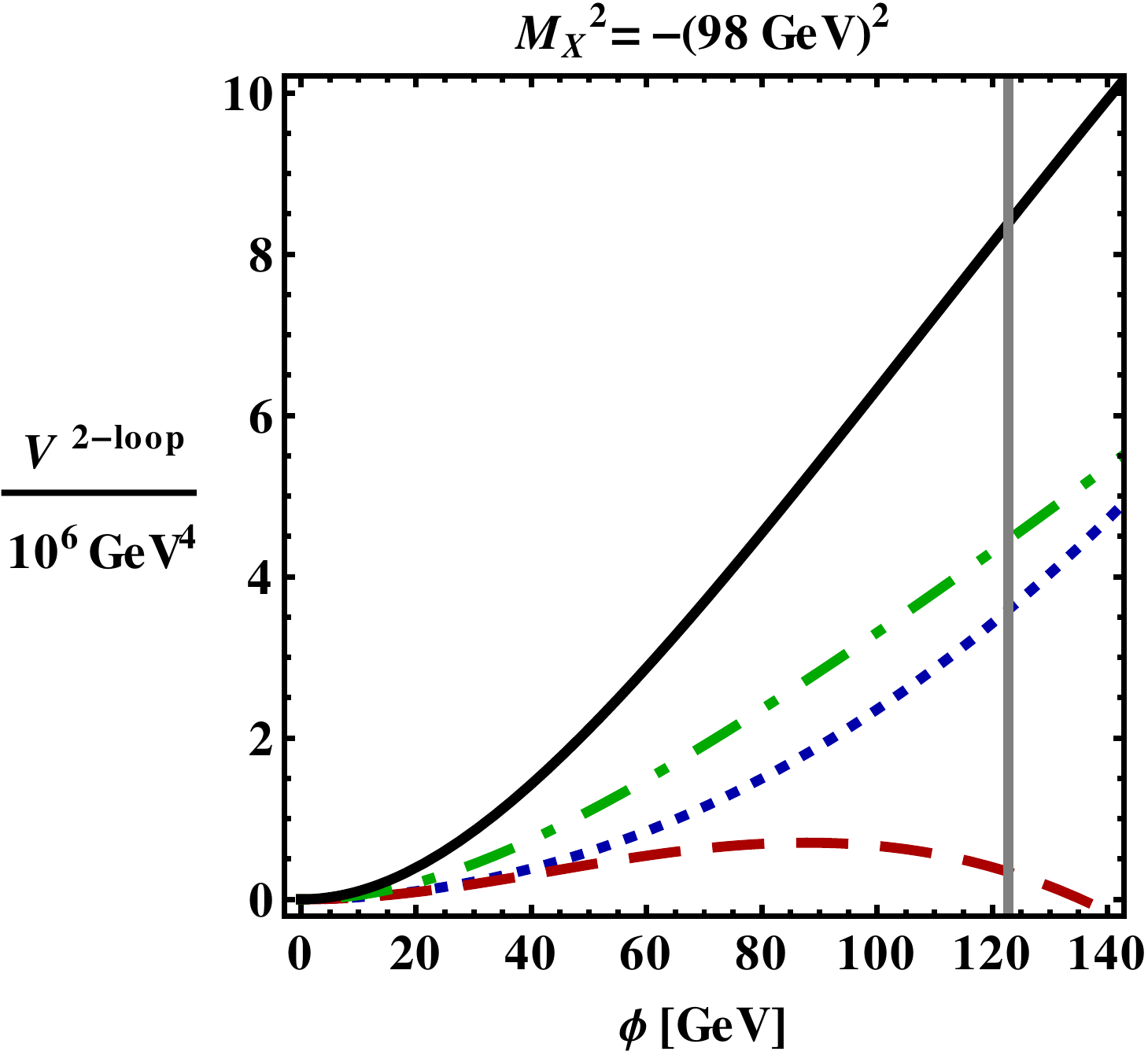}
\end{array}$
\caption{\label{fig:VariousContributionsToPotential} \small We show various contributions to the 2-loop finite temperature potential for the respective 2-loop value of $T_C$ as a function of $\phi$.  Specifically the curves are the total 2-loop contribution [black, solid]; the contribution from the $X_c$-$g$ loop [green, dot-dashed]; the contribution from the $X_c$-$X_c$-$g$ loop [red, dashed]; and the contribution from the $X_c$-$X_c$-$h$, $X_c$-$h$, $X_c$-$\chi$, $X_c$-$X_c$ and pure Standard Model diagrams summed together [blue, dotted].  The gray vertical line marks $\phi_C$.  $M_X^2 = 0$ for the left figure and $M_X^2 = -(98 \mbox{ GeV})^2$ for right figure.   The other parameters are taken to be $Q = 0.7$, $K = 1.6$, and $m_h = 115$ GeV.  The 2-loop analysis yields $\phi_C = 45.1 \mbox{ GeV}$ and $T_C = 130.9 \mbox{ GeV}$ for $M_X^2 = 0$, and $\phi_C = 122.8 \mbox{ GeV}$ and $T_C = 115.9 \mbox{ GeV}$ for $M_X^2 = -(98 \mbox{ GeV})^2$.}
\end{centering}
\end{figure}

We can better understand the dependence on the mass of $X$ by examining the contributions from the scalar-vector ($X_c$-$g$) and scalar-scalar-vector ($X_c$-$X_c$-$g$) diagrams as a function of $M_X^2$.  In Fig.~\ref{fig:VariousContributionsToPotential}, we have plotted the 2-loop finite temperature potential at $T_C$ as a function of $\phi$ from the following (see Fig.~\ref{fig:2LoopDiagrams} for the Feynman diagrams): the total 2-loop contribution [black, solid]; the contribution from the $X_c$-$g$ loop [green, dot-dashed]; the contribution from the $X_c$-$X_c$-$g$ loop [red, dashed]; and the contribution from the $X_c$-$X_c$-$h$, $X_c$-$h$, $X_c$-$\chi$, $X_c$-$X_c$ and pure Standard Model diagrams summed together [blue, dotted].  The graph on the left is for $M_X^2 = 0$ and the one on the right is for $M_X^2 = -(98 \mbox{ GeV})^2$ and the other parameters are taken to be $Q = 0.7$, $K = 1.6$, and $m_h = 115$ GeV for both.  The gray vertical line marks $\phi_C$ for each case: $M_X^2 = 0 \Rightarrow \phi_C = 45.1 \mbox{ GeV};\,T_C = 130.9 \mbox{ GeV}$ and $M_X^2 = -(98 \mbox{ GeV})^2 \Rightarrow \phi_C = 122.8 \mbox{ GeV};\,T_C = 115.9 \mbox{ GeV}$.  

In both cases, the full 2-loop contribution to the potential (the solid black curve) acts as a positive effective mass for $\phi$, pinning it to the origin and postponing the phase transition \cite{Dine:1992vs}, which decreases $T_C$ below its 1-loop value --- it also works to increase $\phi_C$.  The effect on $\phi_C$ can be understood by recalling that for smaller values of $T_C$, $\Pi_c(T_C)$ also decreases.  This implies that $\overline{M}_X$ is smaller at $\phi=0$, which leads to an increase in the value of $\phi_C$, as mentioned above.

To explore the $M_X^2$ dependence of $\phi_C/T_C$ we can compare the left and right panels of Fig.~\ref{fig:VariousContributionsToPotential}.  We see that the contribution from the diagrams which do not involve color (the blue dotted curve) is roughly the same between these two cases, \emph{i.e.}, it has a weak dependence on $M_X^2$ and $T$.  Examining the diagrams that involve gluons (green dash-dotted and red dashed curves), we see a strong dependence on $M_X^2$.  The $X_c$-$g$ contribution is increased by a non-trivial amount from the $M_X^2 = 0$ case to the $M_X^2 \ll 0$ case and in fact becomes the dominant diagram.  Also, when $M_X^2 = 0$, the $X_c$-$X_c$-$g$ contribution is large and negative, decreasing the total size of the potential by a non-trivial amount.  However, when $M_X^2 \ll 0$, this contribution becomes positive (in the region of interest, $\phi \lsim \phi_C$).  It then works in concert with the other 2-loop effects to help postpone the phase transition, thereby increasing the 1\st orderness.  

\subsection{The Electroweak Phase Transition with New Singlet Scalars}\label{sec:EWPTwithSingletScalar}
To contrast with the case just considered, we briefly discuss the model where $X$ is uncolored, denoted by $X_0$.  To make the comparison more straightforward, we assume there exist six identical $X_0$'s.  This matches the scalar degrees of freedom in the colored case.  The Lagrangian is still given by Eq.~(\ref{eq:XLagrangian}).  We find that in this case the 2-loop corrections can increase (for $Q \lesssim 1$) or decrease (for larger values of $Q\gtrsim 1$) $\phi_C/T_C$ with respect to its 1-loop value by up to $\sim 25 \%$.

\begin{figure}[t]
\begin{centering}
\includegraphics[width=.7\textwidth]{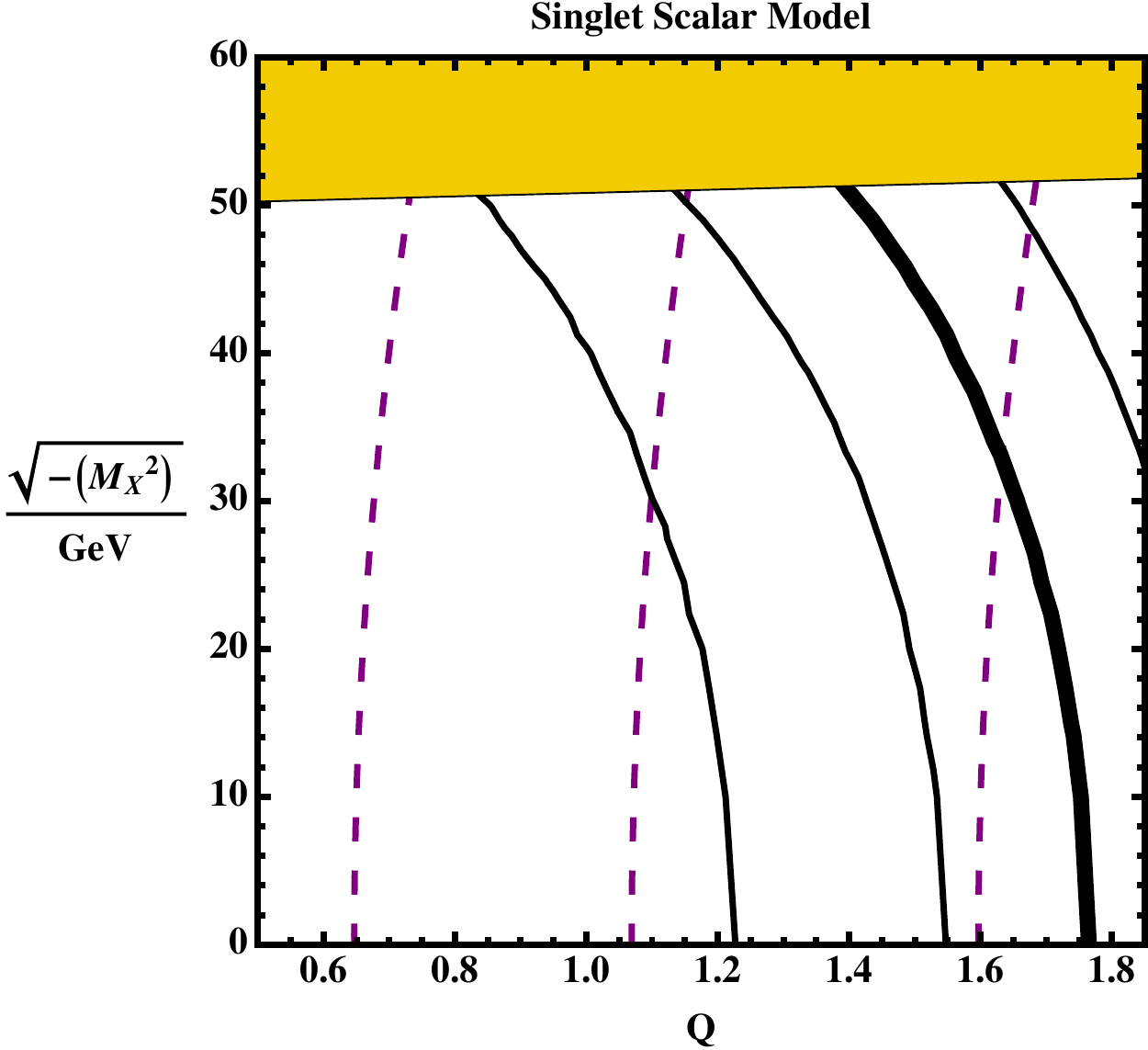}
\caption{\label{fig:ContourMagicNumber2Loop_Singlet} \small Contours of $\phi_C/T_C$ [black, solid] and the physical $X$ mass [purple, dashed] in the $\sqrt{-M_X^2}$ vs.~$Q$ plane in the model with six additional real singlet scalars, $X_0$.  The Higgs mass is fixed to be 115 GeV and $K=1.6$.  The thickest solid contour marks $\phi_C/T_C = 0.9$ and the contours to the right (left) of it mark successive values of $+$ ($-$) 0.2.  The dashed contours are for $X$ masses of 140 GeV, 180 GeV, and 220 GeV from left to right.  In the yellow region there would be a phase transition to a vacuum with $\vev{X_0}\neq 0$ before the electroweak phase transition would occur.  The maximum value of $Q$ plotted corresponds to where the high temperature expansion begins to break down.}
\end{centering}
\end{figure}

In Fig.~\ref{fig:ContourMagicNumber2Loop_Singlet}, we plot the singlet analogue of Fig.~\ref{fig:ContourMagicNumber2Loop_Colored}.  The yellow region is where the transition in the $X_0$ direction would occur before the phase transition for $\phi$, again computed at two loops.\footnote{Here we just impose that the critical temperature in the $\phi$ direction be greater than that in the $X$ direction since we have not preformed the bounce action computation for this model.}  Note, as opposed to the case where $X$ is colored, $\vev{X_0}\neq 0$ could in principle lead to a viable scenario.  However, if $X_0$ acquires a vev, $X_0 - h$ mixing will occur.  This will change the phenomenology of the Higgs sector and the dynamics of the phase transition.  Our goal here is simply to contrast with the model with a colored scalar, a single phase transition, and a Standard Model-like Higgs.  We are thus only interested in the parameter space outside the yellow region.  

Comparing the allowed region in Fig.~\ref{fig:ContourMagicNumber2Loop_Singlet} to the allowed region for the case when $X$ is colored, we see that the size of the viable parameter space has been reduced.  This is largely because the thermal mass for $X$, Eq.~(\ref{eq:PI}), no longer contains a (large) contribution proportional to $g_s^2$ since it is uncolored.  Hence, the requirement that $\vev{\phi}\neq 0$ and $\vev{X_0}=0$ (the solid yellow region) eliminates more parameter space than in the colored case.  There is a mild dependence on the size of $K$.   Specifically, changing $K$ from $1.6$ to $0.8$ while keeping the other parameters fixed implies a change in the strength of the phase transition of order 20\%.

From Fig.~\ref{fig:ContourMagicNumber2Loop_Singlet}, we see that even pushing $M_X^2$ to be as negative as is allowed, requires large values of $Q \simeq 1.4$ to achieve a 1\st order phase transition.  We conclude that while this scenario is viable, perturbativity of the $Q$ coupling imposes an important restriction on the parameter space where a strong phase transition is achieved. 

\section{Collider Phenomenology}\label{sec:ColliderPheno}
If the new scalar driving the phase transition is uncolored, direct discovery of this state is unlikely. It will only be produced via its coupling to the Higgs boson.  So, even if we observe novel CP violation consistent with electroweak baryogenesis (\emph{e.g.}, through an electric dipole moment measurement), we might find ourselves in the uncomfortable scenario of  being unable to determine  the strength of the Higgs phase transition.  However, as demonstrated above, the parameter space of the singlet model is limited.  It seems more natural that if the phase transition is driven by a new scalar, it is colored.  This is good news  --- it opens the possibility of learning about the electroweak phase transition at hadron colliders.  

In this section, we refer to the colored scalars $X_c$ as ``stops" since the present collider phenomenology is essentially identical to this more familiar case.  We focus on two final states which are the simplest to realize and are not yet excluded for stops that are sufficiently  light to effect the 1\st order phase transition we require.  The first is $X_c \rightarrow c\,\chi$, where $c$ is the charm quark and $\chi$ is a new stable neutral state (presumably the dark matter, which we will refer to as a ``neutralino") that we have not discussed, but whose inclusion in a larger model is straightforward.  Its presence need not impact the phase transition.  This is the case studied most often in the literature.  There is additionally an interesting correlation with the dark matter relic abundance since it is possible to be in the ``stop-coannihilation" region \cite{Balazs:2004bu}.  It is also possible that the stop can decay to a pair of jets, $X_c \rightarrow jj$.  In the MSSM analog, this can be realized by including the $R$-parity violating operator $U_{3}^c\,D_i^c\,D_j^c$ in the superpotential.  As described above, for simplicity we have assumed that the $X_c$ is in the fundamental representation of $SU(3)$ for the purposes of the phase transition calculation.  This will have an impact on the collider phenomenology of the scalars as well.

At the Tevatron, a search for $X_c \rightarrow c\, \chi$ has been performed by CDF for 2.6 $\mbox{fb}^{-1}$ \cite{CDFNote9834} and at D$\slashed{0}$ for 1 $\mbox{fb}^{-1}$ \cite{Abazov20081}, resulting in an exclusion contour in the $m_\mr{stop}-m_{\chi}$ plane.  The search is least sensitive when $X_{c}$ and $\chi$ are nearly degenerate.  For example, stops with a mass as low as 96 GeV (at the edge of the LEP bound) are still allowed for a neutralino mass of $\sim 60$ GeV.  If the stop decays to a pair of jets, it is unclear whether any bound from the Tevatron obtains.  However, a search for $X_c X_c \rightarrow (jj)(jj)$ might be possible \cite{Kilic:2008pm}.  In the case where one of the final state jets is a $b$ quark, the situation is potentially more optimistic \cite{Choudhury:2005dg}.  

At the LHC, it may be possible to close the window where there is a small splitting between the stop and the neutralino.  The authors of \cite{Carena:2008mj} suggest searching for stops produced with an additional hard jet.  This jets + $\slashed{E}{}_T$ approach would give a 5$\sigma$ discovery reach at LHC14 for the stop masses up to $\sim 180$ GeV for any value of the neutralino mass.  The larger stop-neutralino splittings only require $\sim 30 \mbox{ fb}^{-1}$ of data while the most degenerate cases require as much as $300 \mbox{ fb}^{-1}$.  Also, the analysis of \cite{Bornhauser:2010mw} suggests stops produced in association with a pair of $b$ quarks can be discovered with 5$\sigma$ significance out to masses of $\sim 270$ GeV with 100 fb$^{-1}$ at LHC14.  Another (albeit more model dependent) signature was proposed in \cite{Kraml:2005kb, Kraml:2006ca}: same sign stops produced from gluinos which each decay to a stop-top pair.  Using 30 fb$^{-1}$ of data (again at LHC14), they demonstrate (fixing $m_{\tilde{g}}=900$ GeV) that a stop with a mass in the entire range of interest to electroweak baryogenesis can be discovered with between 4-5$\sigma$ significance.  This strategy would not apply for the simplest model which includes only one new scalar --- considerations of the phase transition alone do not ensure the presence of gluino-like states.

Given that the LHC7 has already collected and analyzed $\sim 1 \mbox{ fb}^{-1}$ of data, one can wonder if it can already begin to exclude some of the open parameter space where $X_c\rightarrow c\, \chi$.  We simulated events of this type with Madgraph version 4.4.32 \cite{Alwall:2007st} and analyzed the data with PGS4 \cite{Conway} with a modification to implement the anti-$k_t$ jet algorithm.  Using the ``HighPT" cuts of an ATLAS search for one hard jet with missing energy \cite{ATLAS-CONF-2011-096}, we find that there would be $\simeq 60$ events after cuts for 1 fb$^{-1}$ of data.  This should be compared with their background prediction after cuts of $1010\pm 37 \pm 65$ events.  We see that our model would not be observable yet.  The LHC7 may begin to probe the open parameter space with only a modest increase in luminosity.  Furthermore, it is plausible that by tuning the cuts provided in the ATLAS study one could do better than our simple estimate \cite{Chen:2010kq}. Finally, as shown in \cite{Izaguirre:2010nj} it may actually be advantageous to look at 2 jet plus missing energy signatures.  

If instead the dominant stop decay is via the $X_c \rightarrow jj$ channel, the relevant LHC search will be for two di-jet resonances \cite{Kilic:2008ub}.  A search for a pair of color-octet scalars which each decay to a pair of jets was performed at ATLAS \cite{Zhu}.  It excludes a putative scalar gluon partner (sgluon) \cite{Hall:1990hq} with a 1 nb (350 pb) cross section for masses of 100 GeV (190 GeV).  Taking into account the production cross section, this eliminates essentially all slguon masses below 180 GeV.   On the other hand, a scalar in the fundamental representation of SU(3) has a substantially smaller cross section, and no bound applies at the moment.  While this search used only 34 pb$^{-1}$, it may prove to be non-trivial to apply this analysis to the larger data set due to the differences in the 4-jet trigger thresholds \cite{Zerwas}.  We leave a detailed study of the potential to exclude light colored states at the LHC for future work.

\section{Conclusions}\label{sec:Conclusions}
In this paper we have explored the electroweak phase transition in the class of models with one light Higgs boson which has a quartic coupling to a new colored scalar.  We showed that the 2-loop effects could increase the strength of the phase transition by as much as a factor of 3.5 for negative values of the bare mass of a colored scalar.  We also argued that this increase was tied to the color quantum numbers of the scalar and explored these effects by contrasting with a model where new singlet scalars couple to the Higgs.

As we have emphasized, the 2-loop effects are crucial to achieving a 1\st order phase transition in these models.  If a new light scalar is observed, this motivates more precise computations.  In particular, an investigation of how 3-loop effects alter the strength of the electroweak phase transition would be warranted.

From the point of view of discovery, a nightmare scenario does exist, \emph{i.e.}, the singlet scalar model with large quartics.  However, we have argued that the parameter space is limited in this case --- it requires both a large number of scalars, and a very large quartic coupling.  The relative ease of generating a 1\st order phase transition with a colored scalar suggests a promising LHC phenomenology.  Since the colored scalar must be light to avoid Boltzmann suppressed at temperatures relevant to the phase transition, it should be observable in either the jets + $\slashed{E}{}_T$ channel or as a pair of di-jet resonances for the entire mass range of interest.  Care with jet threshold triggers will be crucial for the discovery of the di-jet resonances.

\vspace{-0.2cm}
\subsection*{Acknowledgments}
\vspace{-0.3cm}
\noindent
We want to especially thank Marcela Carena, Germano Nardini, Mariano Quiros, and Carlos Wagner for detailed discussions of their previous work.  This paper also benefited from discussions with Dan Feldman, Can Kilic, Jay Wacker, and Dirk Zerwas.  The work of T.C.~was supported in part by DOE Grants \#DE-FG02-95ER-40899 and \#DE-AC02-76SF00515 and NSF CAREER Grant \#NSF-PHY-0743315.  The work of A.P.~was supported in part by NSF CAREER Grant \#NSF-PHY-0743315 and by DOE Grant \#DE-FG02-95ER40899. 

\bibliography{EWPT}{}
\bibliographystyle{utphys}

\end{document}